\documentclass[twocolumn,times]{aastex63}

\usepackage{amssymb}
\usepackage{graphicx}
\usepackage{makecell}
\usepackage[ruled]{algorithm2e}
\usepackage{amsmath}
\usepackage{amsthm}
\usepackage{booktabs}
\usepackage{lineno}
\usepackage{subfigure}
\usepackage{threeparttable}
\usepackage{hyperref}
\usepackage[squaren]{SIunits}
\usepackage{xspace}
\usepackage{comment}
\usepackage{xcolor}
\usepackage{multirow}
\usepackage{soul, color, xcolor}
\soulregister{\cite}7 % 注册\cite命令
\soulregister{\citep}7 % 注册\citep命令
\soulregister{\citet}7 % 注册\citet命令
\soulregister{\ref}7 % 注册\ref命令
\soulregister{\pageref}7 % 注册\pageref命令
\sethlcolor{yellow} %可以更换
\def\ds@onecolumn{\@twocolumnfalse}
\usepackage{array}
\setwatermarkfontsize{1.5in}

\shortauthors{Xiao ET AL.}
\begin{document}

\title{The Minimum Variation Timescales of X-ray bursts from SGR J1935+2154}
\author{Shuo Xiao*}
\affil{Guizhou Provincial Key Laboratory of Radio Astronomy and Data Processing, Guizhou Normal University, Guiyang 550001, People’s Republic of China}
\affil{School of Physics and Electronic Science, Guizhou Normal University, Guiyang 550001, People’s Republic of China;\\ xiaoshuo@gznu.edu.cn, aijdong@gznu.edu.cn}

\author{Jiao-Jiao Yang}
\affil{Guizhou Provincial Key Laboratory of Radio Astronomy and Data Processing, Guizhou Normal University, Guiyang 550001, People’s Republic of China}
\affil{School of Physics and Electronic Science, Guizhou Normal University, Guiyang 550001, People’s Republic of China;\\ xiaoshuo@gznu.edu.cn, aijdong@gznu.edu.cn}

\author{Xi-Hong Luo}
\affil{Guizhou Provincial Key Laboratory of Radio Astronomy and Data Processing, Guizhou Normal University, Guiyang 550001, People’s Republic of China}
\affil{School of Physics and Electronic Science, Guizhou Normal University, Guiyang 550001, People’s Republic of China;\\ xiaoshuo@gznu.edu.cn, aijdong@gznu.edu.cn}

\author{Shao-Lin Xiong*}
\affil{Key Laboratory of Particle Astrophysics, Institute of High Energy Physics, Chinese Academy of Sciences, Beijing 100049, China;\\ xiongsl@ihep.ac.cn, zhangsn@ihep.ac.cn}

\author{Yuan-Hong Qu}
\affil{Nevada Center for Astrophysics, University of Nevada, Las Vegas, NV 89154}
\affil{Department of Physics and Astronomy, University of Nevada Las Vegas, Las Vegas, NV 89154, USA}

\author{Shuang-Nan Zhang*}
\affil{Key Laboratory of Particle Astrophysics, Institute of High Energy Physics, Chinese Academy of Sciences, Beijing 100049, China;\\ xiongsl@ihep.ac.cn, zhangsn@ihep.ac.cn}
\affil{University of Chinese Academy of Sciences, Chinese Academy of Sciences, Beijing 100049, China}

\author{Wang-Chen Xue}
\affil{Key Laboratory of Particle Astrophysics, Institute of High Energy Physics, Chinese Academy of Sciences, Beijing 100049, China;\\ xiongsl@ihep.ac.cn, zhangsn@ihep.ac.cn}
\affil{University of Chinese Academy of Sciences, Chinese Academy of Sciences, Beijing 100049, China}

\author{Xiao-Bo Li}
\affil{Key Laboratory of Particle Astrophysics, Institute of High Energy Physics, Chinese Academy of Sciences, Beijing 100049, China;\\ xiongsl@ihep.ac.cn, zhangsn@ihep.ac.cn}

\author{You-Li Tuo}
\affil{Institut für Astronomie und Astrophysik, University of Tübingen, Sand 1, 72076 Tübingen, Germany}

\author{Ai-Jun Dong*}
\affil{Guizhou Provincial Key Laboratory of Radio Astronomy and Data Processing, Guizhou Normal University, Guiyang 550001, People’s Republic of China}
\affil{School of Physics and Electronic Science, Guizhou Normal University, Guiyang 550001, People’s Republic of China;\\ xiaoshuo@gznu.edu.cn, aijdong@gznu.edu.cn}

\author{Ru-shuang Zhao}
\affil{Guizhou Provincial Key Laboratory of Radio Astronomy and Data Processing, Guizhou Normal University, Guiyang 550001, People’s Republic of China}
\affil{School of Physics and Electronic Science, Guizhou Normal University, Guiyang 550001, People’s Republic of China;\\ xiaoshuo@gznu.edu.cn, aijdong@gznu.edu.cn}

\author{Shi-Jun Dang}
\affil{Guizhou Provincial Key Laboratory of Radio Astronomy and Data Processing, Guizhou Normal University, Guiyang 550001, People’s Republic of China}
\affil{School of Physics and Electronic Science, Guizhou Normal University, Guiyang 550001, People’s Republic of China;\\ xiaoshuo@gznu.edu.cn, aijdong@gznu.edu.cn}

\author{Lun-Hua Shang}
\affil{Guizhou Provincial Key Laboratory of Radio Astronomy and Data Processing, Guizhou Normal University, Guiyang 550001, People’s Republic of China}
\affil{School of Physics and Electronic Science, Guizhou Normal University, Guiyang 550001, People’s Republic of China;\\ xiaoshuo@gznu.edu.cn, aijdong@gznu.edu.cn}

\author{Qing-Bo Ma}
\affil{Guizhou Provincial Key Laboratory of Radio Astronomy and Data Processing, Guizhou Normal University, Guiyang 550001, People’s Republic of China}
\affil{School of Physics and Electronic Science, Guizhou Normal University, Guiyang 550001, People’s Republic of China;\\ xiaoshuo@gznu.edu.cn, aijdong@gznu.edu.cn}

\author{Ce Cai}
\affil{College of Physics, Hebei Normal University, 20 South Erhuan Road, Shijiazhuang, 050024, China}

\author{Jin Wang}
\affil{Key Laboratory of Particle Astrophysics, Institute of High Energy Physics, Chinese Academy of Sciences, Beijing 100049, China;\\ xiongsl@ihep.ac.cn, zhangsn@ihep.ac.cn}

\author{Ping Wang}
\affil{Key Laboratory of Particle Astrophysics, Institute of High Energy Physics, Chinese Academy of Sciences, Beijing 100049, China;\\ xiongsl@ihep.ac.cn, zhangsn@ihep.ac.cn}

\author{Cheng-Kui Li}
\affil{Key Laboratory of Particle Astrophysics, Institute of High Energy Physics, Chinese Academy of Sciences, Beijing 100049, China;\\ xiongsl@ihep.ac.cn, zhangsn@ihep.ac.cn}

\author{Shu-Xu Yi}
\affil{Key Laboratory of Particle Astrophysics, Institute of High Energy Physics, Chinese Academy of Sciences, Beijing 100049, China;\\ xiongsl@ihep.ac.cn, zhangsn@ihep.ac.cn}

\author{Zhen Zhang}
\affil{Key Laboratory of Particle Astrophysics, Institute of High Energy Physics, Chinese Academy of Sciences, Beijing 100049, China;\\ xiongsl@ihep.ac.cn, zhangsn@ihep.ac.cn}

\author{Ming-Yu Ge}
\affil{Key Laboratory of Particle Astrophysics, Institute of High Energy Physics, Chinese Academy of Sciences, Beijing 100049, China;\\ xiongsl@ihep.ac.cn, zhangsn@ihep.ac.cn}

\author{Shi-Jie Zheng}
\affil{Key Laboratory of Particle Astrophysics, Institute of High Energy Physics, Chinese Academy of Sciences, Beijing 100049, China;\\ xiongsl@ihep.ac.cn, zhangsn@ihep.ac.cn}

\author{Li-Ming Song}
\affil{Key Laboratory of Particle Astrophysics, Institute of High Energy Physics, Chinese Academy of Sciences, Beijing 100049, China;\\ xiongsl@ihep.ac.cn, zhangsn@ihep.ac.cn}

\author{Wen-Xi Peng}
\affil{Key Laboratory of Particle Astrophysics, Institute of High Energy Physics, Chinese Academy of Sciences, Beijing 100049, China;\\ xiongsl@ihep.ac.cn, zhangsn@ihep.ac.cn}

\author{Xiang-Yang Wen}
\affil{Key Laboratory of Particle Astrophysics, Institute of High Energy Physics, Chinese Academy of Sciences, Beijing 100049, China;\\ xiongsl@ihep.ac.cn, zhangsn@ihep.ac.cn}

\author{Xin-Qiao Li}
\affil{Key Laboratory of Particle Astrophysics, Institute of High Energy Physics, Chinese Academy of Sciences, Beijing 100049, China;\\ xiongsl@ihep.ac.cn, zhangsn@ihep.ac.cn}

\author{Zheng-Hua An}
\affil{Key Laboratory of Particle Astrophysics, Institute of High Energy Physics, Chinese Academy of Sciences, Beijing 100049, China;\\ xiongsl@ihep.ac.cn, zhangsn@ihep.ac.cn}

\author{Xin Xu}
\affil{Guizhou Provincial Key Laboratory of Radio Astronomy and Data Processing, Guizhou Normal University, Guiyang 550001, People’s Republic of China}
\affil{School of Physics and Electronic Science, Guizhou Normal University, Guiyang 550001, People’s Republic of China;\\ xiaoshuo@gznu.edu.cn, aijdong@gznu.edu.cn}

\author{Yue Wang}
\affil{Key Laboratory of Particle Astrophysics, Institute of High Energy Physics, Chinese Academy of Sciences, Beijing 100049, China;\\ xiongsl@ihep.ac.cn, zhangsn@ihep.ac.cn}
\affil{University of Chinese Academy of Sciences, Chinese Academy of Sciences, Beijing 100049, China}

\author{Chao Zheng}
\affil{Key Laboratory of Particle Astrophysics, Institute of High Energy Physics, Chinese Academy of Sciences, Beijing 100049, China;\\ xiongsl@ihep.ac.cn, zhangsn@ihep.ac.cn}
\affil{University of Chinese Academy of Sciences, Chinese Academy of Sciences, Beijing 100049, China}

\author{Yan-Qiu Zhang}
\affil{Key Laboratory of Particle Astrophysics, Institute of High Energy Physics, Chinese Academy of Sciences, Beijing 100049, China;\\ xiongsl@ihep.ac.cn, zhangsn@ihep.ac.cn}
\affil{University of Chinese Academy of Sciences, Chinese Academy of Sciences, Beijing 100049, China}

\author{Jia-Cong Liu}
\affil{Key Laboratory of Particle Astrophysics, Institute of High Energy Physics, Chinese Academy of Sciences, Beijing 100049, China;\\ xiongsl@ihep.ac.cn, zhangsn@ihep.ac.cn}
\affil{University of Chinese Academy of Sciences, Chinese Academy of Sciences, Beijing 100049, China}

\author{Bin Zhang}
\affil{Guizhou Provincial Key Laboratory of Radio Astronomy and Data Processing, Guizhou Normal University, Guiyang 550001, People’s Republic of China}
\affil{School of Physics and Electronic Science, Guizhou Normal University, Guiyang 550001, People’s Republic of China;\\ xiaoshuo@gznu.edu.cn, aijdong@gznu.edu.cn}

\author{Wei Xie}
\affil{Guizhou Provincial Key Laboratory of Radio Astronomy and Data Processing, Guizhou Normal University, Guiyang 550001, People’s Republic of China}
\affil{School of Physics and Electronic Science, Guizhou Normal University, Guiyang 550001, People’s Republic of China;\\ xiaoshuo@gznu.edu.cn, aijdong@gznu.edu.cn}

\author{Jian-Chao Feng}
\affil{Guizhou Provincial Key Laboratory of Radio Astronomy and Data Processing, Guizhou Normal University, Guiyang 550001, People’s Republic of China}
\affil{School of Physics and Electronic Science, Guizhou Normal University, Guiyang 550001, People’s Republic of China;\\ xiaoshuo@gznu.edu.cn, aijdong@gznu.edu.cn}

\author{De-Hua Wang}
\affil{Guizhou Provincial Key Laboratory of Radio Astronomy and Data Processing, Guizhou Normal University, Guiyang 550001, People’s Republic of China}
\affil{School of Physics and Electronic Science, Guizhou Normal University, Guiyang 550001, People’s Republic of China;\\ xiaoshuo@gznu.edu.cn, aijdong@gznu.edu.cn}

\author{Qi-Jun Zhi}
\affil{Guizhou Provincial Key Laboratory of Radio Astronomy and Data Processing, Guizhou Normal University, Guiyang 550001, People’s Republic of China}
\affil{School of Physics and Electronic Science, Guizhou Normal University, Guiyang 550001, People’s Republic of China;\\ xiaoshuo@gznu.edu.cn, aijdong@gznu.edu.cn}

% \author{et al.}

%最小光变和能量关系

\begin{abstract}
The minimum variation timescale (MVT) of soft gamma-ray repeaters can be an important probe to estimate the emission region in pulsar-like models, as well as the Lorentz factor and radius of the possible relativistic jet in gamma-ray
burst (GRB)-like models, thus revealing their progenitors and physical mechanisms. In this work, we systematically study the MVTs of hundreds of X-ray bursts (XRBs) from SGR J1935+2154 observed by {\it Insight}-HXMT, GECAM and Fermi/GBM from July 2014 to Jan 2022 through the Bayesian Block algorithm. We find that the MVTs peak at $\sim$ 2 ms, corresponding to a light travel time size of about 600 km, which supports the magnetospheric origin in pulsar-like models. The shock radius and the Lorentz factor of the jet are also constrained in GRB-like models. Interestingly, the MVT of the XRB associated with FRB 200428 is $\sim$ 70 ms, which is longer than that of most bursts and implies its special radiation mechanism.
Besides, the median of MVTs is 7 ms, shorter than the median MVTs of 40 ms and 480 ms for short GRBs or long GRBs, respectively. However, the MVT is independent of duration, similar to GRBs.
Finally, we investigate the energy dependence of MVT and suggest that there is a marginal evidence for a power-law relationship like GRBs but the rate of variation is at least about an order of magnitude smaller. These features may provide an approach to identify bursts with a magnetar origin.

\end{abstract}


\section{Introduction}
Timing analysis is one of the most important approaches to reveal the physical properties of astronomical sources. For example, the traditional duration measures such as T90 and T100, which are the periods during which the total
90\% or 100\% of counts are received. Gamma-ray
bursts (GRBs) can be divided into long GRBs (LGRBs) and short GRBs (SGRBs) according to their T90, which are widely considered to originate from massive star core collapses and binary neutron star or neutron star-black hole mergers (\citealp{woosley2006supernova}; \citealp{abbott2017gravitational}), respectively. Similarly, soft gamma-ray repeaters (SGRs) can be divided into three kinds of bursts according to their duration \citep{2006csxs.book..547W}: short-duration bursts, intermediate bursts (IBs) and giant flares (GFs). Besides their total released energy, their shapes of light curves are also different, for example, a GF usually has a hard initial spike, which is followed by rapidly decaying tail (e.g. \citealp{hurley1999giant}; \citealp{yang2020grb}; \citealp{svinkin2021bright}). However, the duration only describes the total emission properties of a burst, which does not capture the information concerning individual pulses in a burst.
Therefore, several approaches have been utilized to characterize the timescales for astrophysical sources. For example, \cite{guidorzi2016individual} define the GRB dominant variation timescale as $\tau=1/f_{b}$, where $f_b$ is the break frequency as determined by the bent power-law model in the power density spectra (PDS). Therefore, if many peaks (or pulses) in a burst more likely have a specific timescale, the resulting PDS has a dominant timescale. 

On the other hand, the minimum variation timescale (MVT) not only can give insight into the dynamics of the energy dissipation and the dissipation region \citep{1978Natur.271..525S, fenimore1993escape, 2007ApJ...660..556T,ackermann2014fermi, golkhou2014uncovering,golkhou2015energy}, but also distinguish the origin of a burst (e.g. \citealp{yang2020grb,xiao2022quasi}). \cite{2013MNRAS.432..857M} defined the MVT $\Delta t_{\rm mvt}$ in a light curve of a GRB as the transition from a time-scaling region to that of white noise in frequency domain based on the Haar Wavelet analysis that is similar to the Fourier analysis, and they reported a strong positive correlation between MVT and the rise time of the shortest pulse in GRB but the former is smaller \cite{maclachlan2012minimum}. However, the approach does not measure an intrinsic MVT but rather the shortest observable timescale, which tends to heavily depend on the signal-to-noise of the data. Therefore, \cite{golkhou2014uncovering,golkhou2015energy} defined the MVT as the transition between correlated and uncorrelated variability, which is approximately the same as the rise time of the shortest pulse in a light curve. Recently, \cite{2023A&A...671A.112C} defined MVT as the full width half maximum duration of the shortest pulse in time domain and reported a very strong correlation between their results and \cite{golkhou2014uncovering}, but they are not completely consistent, in particular, some differ by more than a decade. Besides, Some authors found that half of the shortest Bayesian block (BB) is consistent with the MVT obtained by the Wavelet analysis of individual GRBs (e.g. \citealp{ackermann2014fermi,vianello2018bright}), but to our knowledge, the MVT estimated by BB has not been systematically compared with the results by other methods to prove its robustness, which will be demonstrated in this work.

SGR J1935+2154 is a well-known Galactic magnetar, which is currently the only one that has been observed with a fast radio burst (FRB) associated with an X-ray burst (XRB), indicating that at least some FRBs originate from magnetar activities (e.g. \citealp{li2021hxmt,bochenek2020fast,mereghetti2020integral,younes2021broadband,ridnaia2021peculiar,tavani2021x}). The spin period and spin-down rate are 3.24 s and $1.43(1)\times10^{-11} {\rm s\ s^{-1}}$, thus the dipole-magnetic field can be estimated as $2.2\times10^{14}$ G \citep{israel2016discovery}. \cite{xiao2023discovery} recently reported that the XRBs from it have a linear energy dependence of the spectral lag similar to Crab pulsar. The physical mechanisms responsible for FRBs and XRBs from magnetars can mainly be divided into pulsar-like models and GRB-like models (see \citealp{2020Natur.587...45Z,2022arXiv221203972Z} for review); the study of MVT of magnetar XRBs may be used to estimate the emission region in pulsar-like models, as well as Lorentz factor and radius of the possible relativistic jet in GRB-like models.
Besides, the magnetar is very active and we therefore have a large sample, which has about 700 bursts from it observed by {\it Insight}-HXMT \citep{li2021hxmt,cai2022insight}, GECAM \citep{xiao2022energetic,xie2022revisit} and Fermi/GBM \citep{lin2020fermi,zou2021periodicity} from July 2014 to Jan 2022. These three satellites have high temporal resolution of $2\ {\rm \mu s}$ \citep{liu2020high,xiao2020deadtime}, $0.1\ {\rm \mu s}$ \citep{xiao2022ground} and $2\ {\rm \mu s}$ \citep{meegan2009fermi} as well as small deadtime for a normal event of $\sim 4\ {\rm \mu s}$ \citep{xiao2020deadtime}, $4\ {\rm \mu s}$ \citep{liu2021sipm} and $2.6\ {\rm \mu s}$ \citep{meegan2009fermi}, respectively, which are beneficial for MVT analyses.

However, to our knowledge, the MVTs of the XRBs from SGR J1935+2154 have not been systematically studied, which is the focus of this work. We present our sample selection and the method of estimating MVT using BB in Section 2, and give the results of the distribution, as well as the duration and energy dependence of MVT, and comparison with GRBs in Section 3. Finally, discussion and summary are given in Section 4.

\begin{figure}
\centering
\includegraphics[width=\columnwidth]{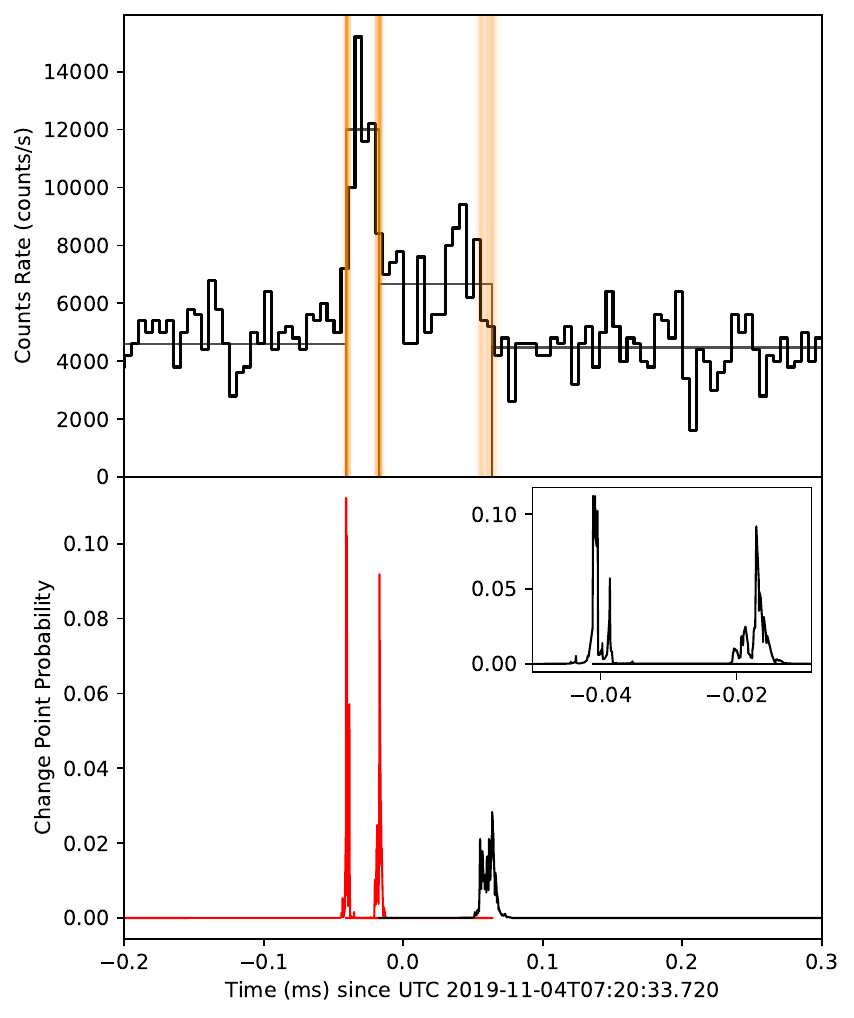}
\caption{An illustration of MVT calculation. Top panel: the 5 ms binned light curve in 8-100 keV of a relatively weak burst ($T_0$=UTC 2019-11-04T07:20:33.720) from SGR J1935+2154 observed by the GBM NaI \#0, 1, 2, 9 and 10, the orange vertical lines are the locations of the change points obtained by the Bayesian block algorithm, and the color depth represents the probability. Bottom panel: approximate probability distribution for the locations of each change point. By calculating the 68\% confidence interval between two shortest time interval change points (red lines), the estimated MVT obtained is $12.0^{+2.3}_{-1.9}$ ms, which is approximated as half the duration of the shortest pulse, that is is from T0-$41.1^{+2.1}_{-0.1}$ ms to T0-$17.1^{+1.0}_{-1.9}$ ms in the light curve.}\label{lc_19110430593}
\end{figure}

\begin{figure}
\centering
\includegraphics[width=\columnwidth]{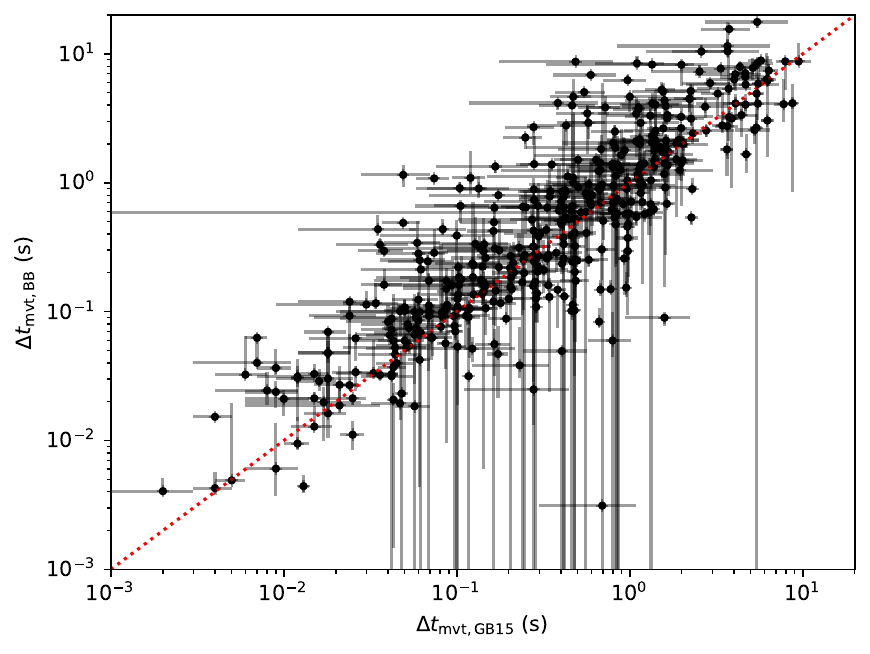}
\caption{Comparison between the MVTs estimated by BB (this work) and Haar Wavelet analysis \citep{golkhou2015energy} for a common sample of Fermi/GBM GRBs, the latter results are from \cite{golkhou2015energy}. The red dashed lines are the equivalent line.}\label{GB15_BB_va}
\end{figure}

\begin{figure*}
\centering
\begin{minipage}[t]{0.46\textwidth}
\centering
\includegraphics[width=\columnwidth]{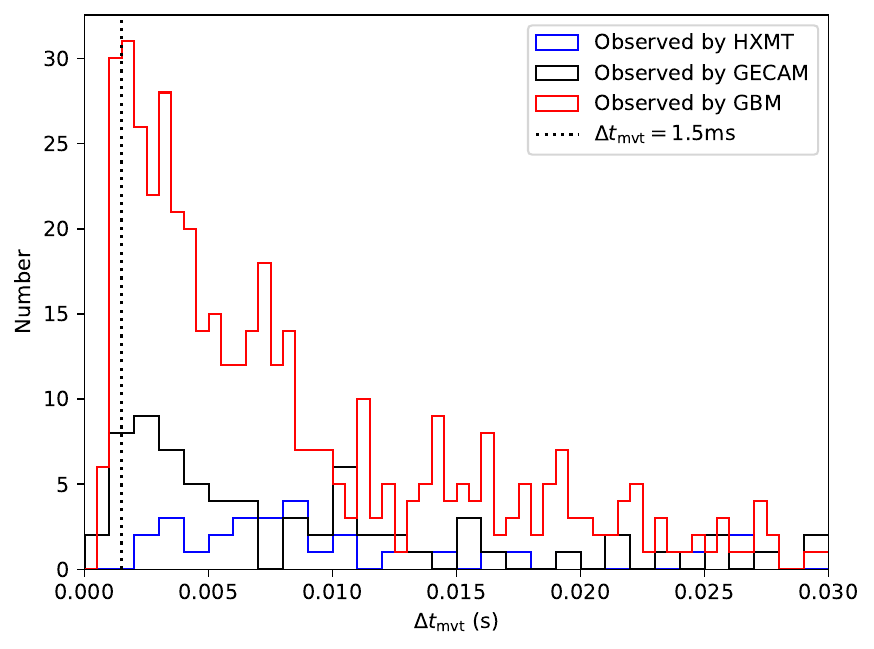}
\end{minipage}
\begin{minipage}[t]{0.46\textwidth}
\centering
\includegraphics[width=\columnwidth]{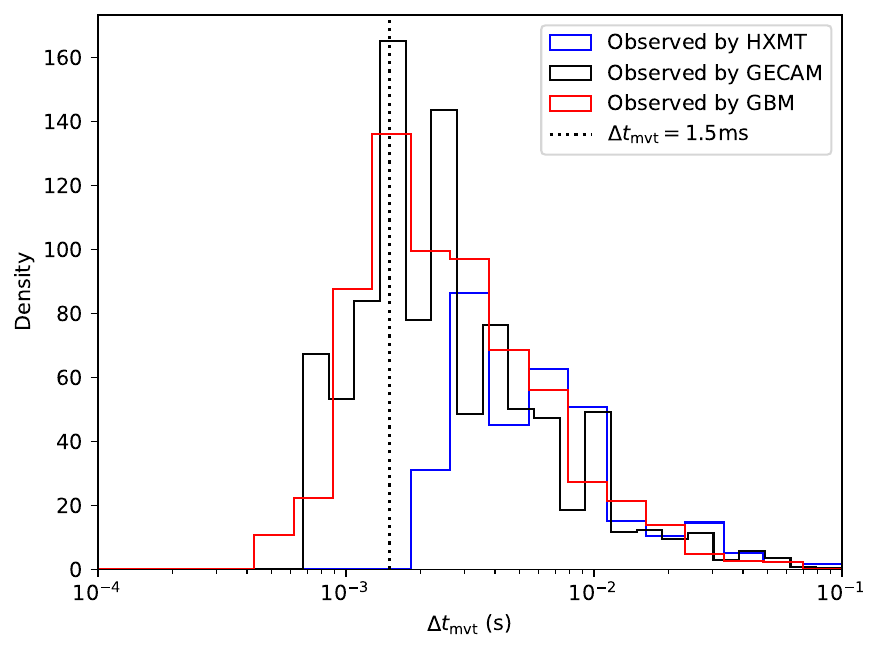}
\end{minipage}
\caption{Distribution of the MVTs ($\Delta t_{\rm mvt}$) for bursts from SGR J1935+2154 observed by GBM, GECAM and HXMT, respectively. The horizontal coordinates of the left panel and right panel are linearly uniform and logarithmically uniform, respectively. The vertical coordinate of the right panel display the bin's count divided by the total number of counts and the bin width (i.e. density).}\label{tmins}
\end{figure*}

\begin{figure*}
\centering
\begin{minipage}[t]{0.48\textwidth}
\centering
\includegraphics[width=\columnwidth]{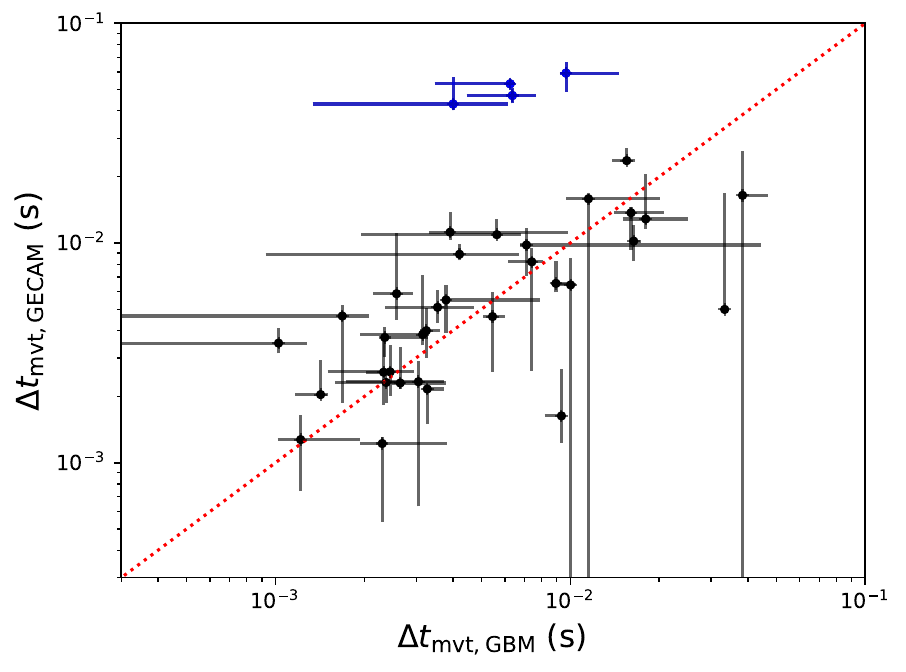}
\end{minipage}
\begin{minipage}[t]{0.48\textwidth}
\centering
\includegraphics[width=\columnwidth]{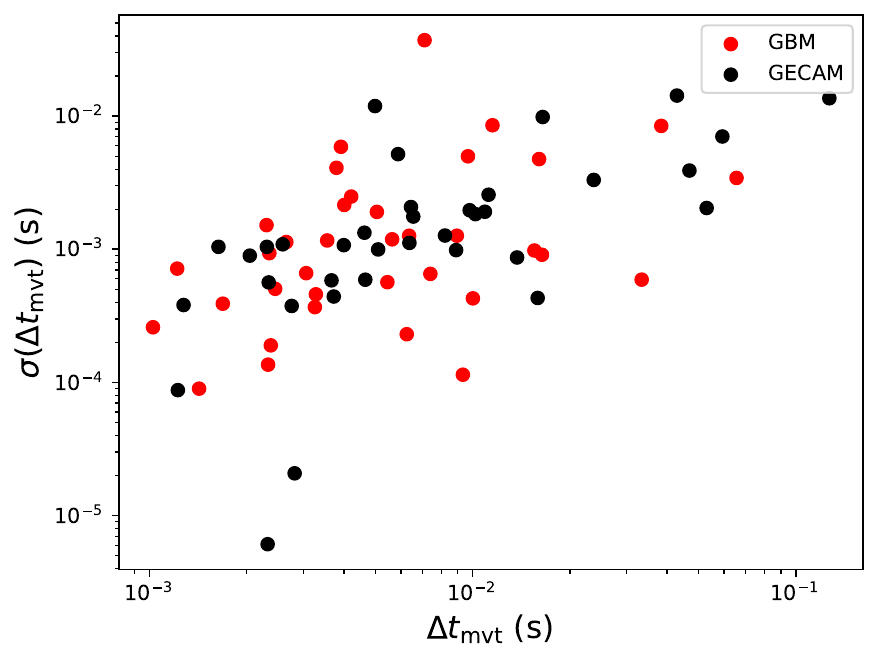}
\end{minipage}
% \begin{minipage}[t]{0.325\textwidth}
% \centering
% \includegraphics[width=\columnwidth]{cross_gbmgecam_e_e.pdf}
% \end{minipage}
\caption{Left panel: consistency check of the MVTs measured by GBM and GECAM, the four blue error bars represent bursts (UTCs 2021-09-12T16:52:07.950, 2021-09-18T22:58:52.150, 2022-01-05T07:06:40.800 and 2022-01-12T01:03:46.900) that deviate obviously from the equivalent line (i.e. the red dashed line), we find that they have significantly lower signal-to-noise ratios on the GECAM than on the GBM, resulting in shorter pulses that are not observed by the former. Right panel: the error in MVT vs. MVT.}\label{gbm_gecam}
\end{figure*}

\begin{figure*}
\centering
\begin{minipage}[t]{0.48\textwidth}
\centering
\includegraphics[width=\columnwidth]{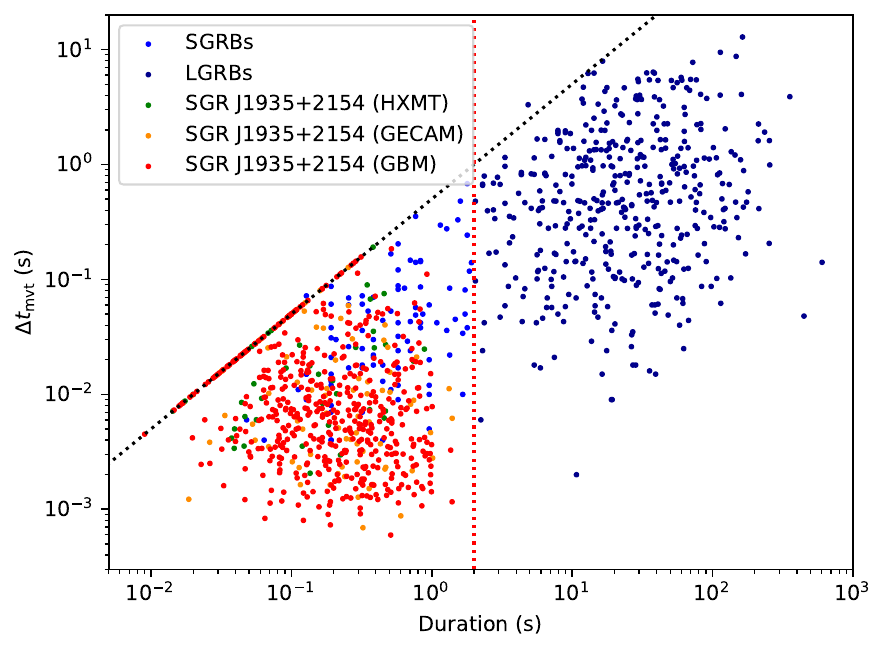}
\end{minipage}
\begin{minipage}[t]{0.47\textwidth}
\centering
\includegraphics[width=\columnwidth]{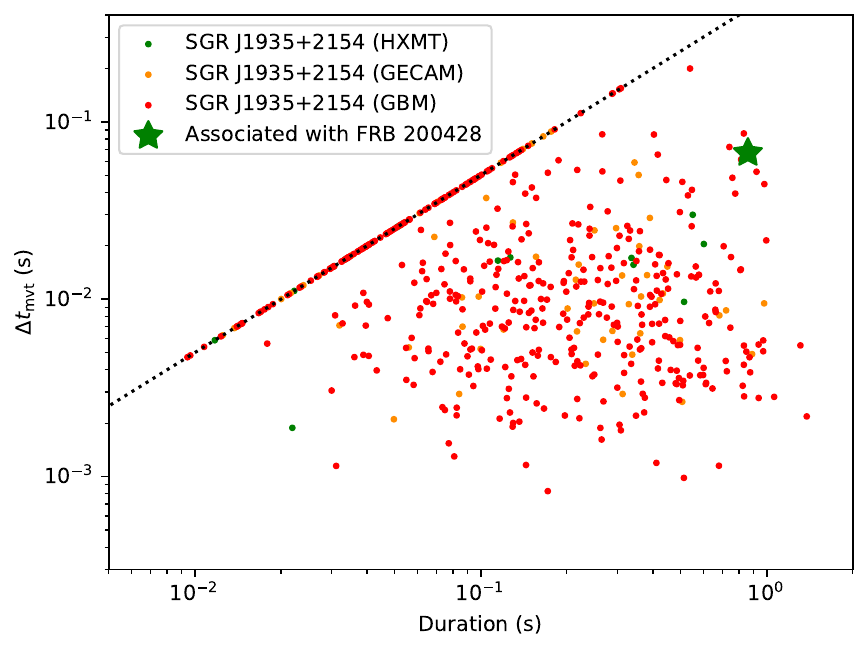}
\end{minipage}
\caption{Left panel: MVT ($\Delta t_{\rm mvt}$) vs. the duration ($T_{\rm bb}$ or T90) for SGR J1935+2154 (red, yellow and green dots), SGRB (blue dots) and LGRB (dark blue dots) in the full energy bands (i.e. 8-100 keV for SGR J1935+2154 and 8-1000 keV for GRB). The red dashed line is the 2 s that is usually considered the division between SGRB and LGRB, the black dashed line represents MVT equal to half the duration, which implies that the burst is single pulse. The MVTs of GRBs are from \cite{golkhou2015energy}. Right panel: MVT ($\Delta t_{\rm mvt}$) vs. the duration ($T_{\rm bb}$ or T90) for SGR J1935+2154 in 15-25 keV, the green star represents the burst associated with FRB 200428 observed by HXMT/ME.}\label{3sources_mvt_t}
\end{figure*}

\begin{figure}
\centering
\includegraphics[width=\columnwidth]{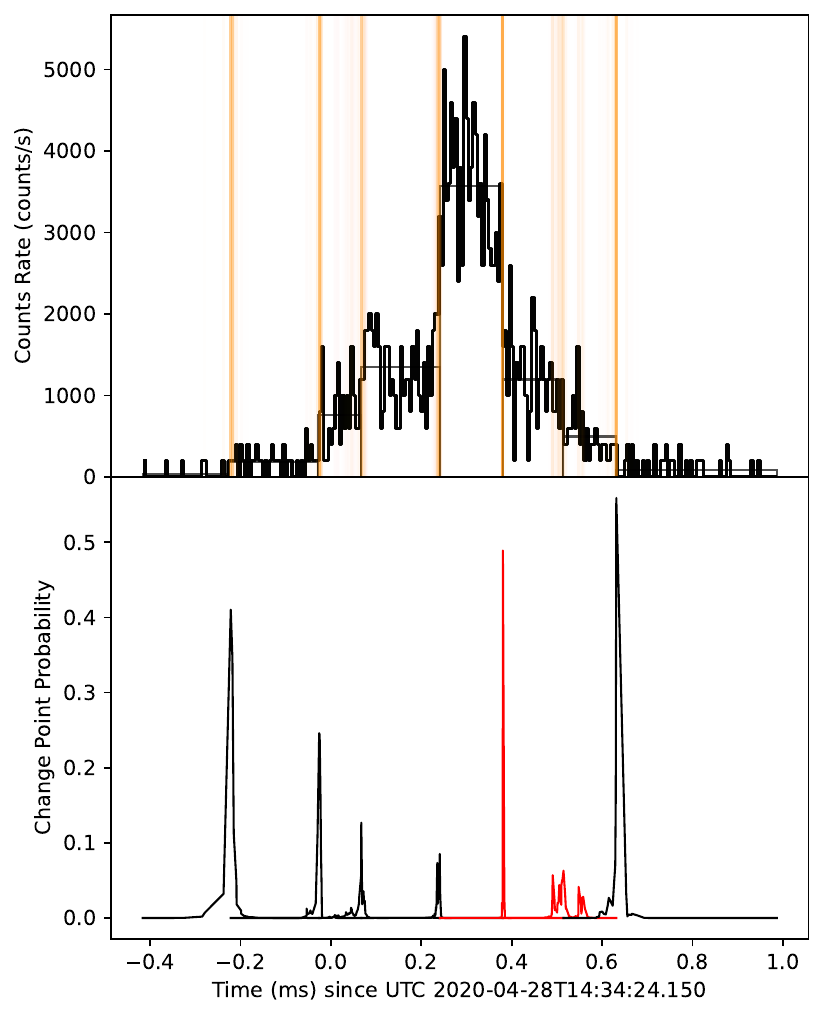}
\caption{Top panel: the 5 ms binned light curve in 15-25 keV of the burst ($T_0$=UTC 2020-04-28T14:34:24.150) associated with FRB 200428 from SGR J1935+2154 observed by the HXMT/ME. Bottom panel: approximate probability distribution for the locations of each change point. By calculating the 68\% confidence interval between two shortest time interval change points (red lines), the estimated MVT obtained is $66.8^{+33.9}_{-23.4}$ ms, which is approximated as half the duration of the shortest pulse, that is from T0+$380.3^{+0.1}_{-0.3}$ ms to T0+$514.0^{+33.8}_{-23.4}$ ms in the light curve. Note that the MVT obtained for the fourth and fifth change points is consistent with the above result within the error. But the shortest block between the second and third change points overlap with probability $2\times10^{-5}$, which is greater than the threshold $3\times10^{-7}$.}\label{FRB_LC}
\end{figure}

\section{SAMPLE SELECTION and Methodology}

\subsection{Data selection, saturation and “timing glitches” screening}
We collect the bursts from SGR J1935+2154 observed by {\it Insight}-HXMT, GECAM and Fermi/GBM from July 2014 to Jan 2022 \citep{lin2020fermi,li2021hxmt,zou2021periodicity,xiao2022energetic,cai2022insight,xie2022revisit,xiao2023discovery}. Note that, since GECAM was launched in December 2020, its bursts are collected after this date. Besides, the bursts for HXMT we selected are only from April 28 2020 to May 24 2020 during the month-long dedicated pointing observation \citep{cai2022insight}, and the bursts that saturated the HXMT detectors due to their extreme brightness of above $\sim$ 30000 counts/s \citep{xiao2020deadtime} are filtered out. For Fermi/GBM, when the summed count rate of all detectors exceeds $\sim$ 375000 counts/s data rate limit of the high speed science data bus, the majority of TTE data are unrecoverable due to the loss of TTE telemetry packet \citep{meegan2009fermi}. Fortunately, thanks to the optimized design of GECAM \citep{liu2021sipm}, to date the bursts from SGR J1935+2154 did not suffer any obvious saturation.

On the other hand, we need to be careful about a data issue “timing glitches”, which is a known GBM hardware anomaly (e.g. dips and peaks in a light curve). These phenomena are understood and have been studied since they are found by \cite{briggs2013terrestrial}. In this work, we check for saturation and “timing glitches” using event-by-event (TTE or Evt) data based on the Bayesian blocks method, i.e., whether blocks with a count of 0 or very low occur, or count rate is about twice the adjacent count rate. After removing these bursts, a total of 41, 83 and 504 bursts from SGR J1935+2154 are collected for HXMT, GECAM, and GBM, respectively.

\subsection{Selection of energy range and detectors}
The default energy range used in our analysis is 8-100 keV, considering the energy spectrum of the bursts and the energy response of the three satellites. For {\it Insight}-HXMT, which includes the High Energy X-ray telescope (HE) with 20-250 keV \citep{liu2020high}, the Medium Energy X-ray
telescope (ME) with 5-30 keV \citep{cao2020medium} and the Low Energy X-ray telescope
(LE) with 1-15 keV \citep{chen2020low}, we select 8-30 and 30-100 keV for ME and HE according to their effective areas \citep{zhang2020overview}, respectively. Note that the burst in association with FRB 200428 suffered significant saturation in HE, thus only the ME data are used.
For GECAM and GBM, those detectors of GECAM and GBM with the incident angle of SGR J1935+2154 less than 60 degrees are selected to improve the statistics. Note that the energy threshold of GECAM may in fact be greater than 8 keV due to the instrumental effects. 

In addition, to investigate the energy dependence of the MVT, we also calculate the MVT for energy ranges 8-15, 15-25, 25-40 and 40-100 keV, respectively.

\subsection{MVT calculation}

Bayesian Block algorithm is a powerful tool of nonparametric analysis for the time series data to identify and characterize statistically significant variations (See \citealp{scargle2013studies} for details). In this work, we use the EvT/TTE data to get the best change points (or  Bayesian block representations), and then define half the length of the two change points with the shortest time interval (or shortest block) as MVT. The false positive rate $p_0$ of reporting detection of a change point is set to 0.05.

We can estimate the approximate probability of each change point at different locations by fixing all but that change point (See \citealp{scargle2013studies} for details). Thus, the probability of overlap between two adjacent change points can be calculated, and we conservatively  must be less than the threshold $3\times10^{-7}$ (corresponding to a Gaussian-equivalent significance of 5$\sigma$) to be considered a valid MVT $\Delta t_{\rm mvt}$.

To estimate the uncertainty in $\Delta t_{\rm mvt}$, the 68\% confidence interval of each change point is obtained based on its probability at different locations, and then the uncertainty of MVT can be obtained by the propagation of errors.

As an illustration, Fig.~\ref{lc_19110430593} shows the light curve and MVT calculation of a relatively weak burst ($T_0$=UTC 2019-11-04T07:20:33.720) from SGR J1935+2154 observed by GBM. The two change points with the shortest interval are T0-$41.1^{+2.1}_{-0.1}$ ms and T0-$17.1^{+1.0}_{-1.9}$ ms, the probability of overlap between them is $8\times10^{-15}$, which is less than the threshold $3\times10^{-7}$. Then the estimated MVT obtained is $12.0^{+2.3}_{-1.9}$ ms, we can find it is approximated as half the duration of the shortest pulse in the burst.

\subsection{Comparison of the results by Bayesian Block and Haar wavelet method in \cite{golkhou2015energy}}
To verify the validity of calculating MVT by Bayesian Block algorithm, we calculate the MVT of the GRBs in \cite{golkhou2015energy}, and compare them with their results calculated by the Haar Wavelet analysis. The comparison between $\Delta t_{\rm mvt\_{BB}}$ and the corresponding $\Delta t_{\rm mvt\_{GB15}}$ estimated by \cite{golkhou2015energy} is shown in Fig.~\ref{GB15_BB_va}, the results obtained by the two different method evidently correlate over four decades and some scatter around equivalent line, which means they have a strong correlation. We simply fit by least squares method to obtain $\Delta t_{\rm mvt\_{BB}}=\sim 1.2\times t_{\rm mvt\_{GB15}}$, which suggests that their results are very similar but not identical, this is due to the fact that the BB method approximates the MVT as half the duration of the shortest pulse, however the MVT obtained by Haar Wavelet analysis approximates the rise time \cite{golkhou2015energy}. For GRB, usually the pulse rise time is shorter than the fall time, thus this may be one of the reasons for our slightly larger results.

\begin{figure}
\centering
\begin{minipage}[t]{0.46\textwidth}
\centering
\includegraphics[width=\columnwidth]{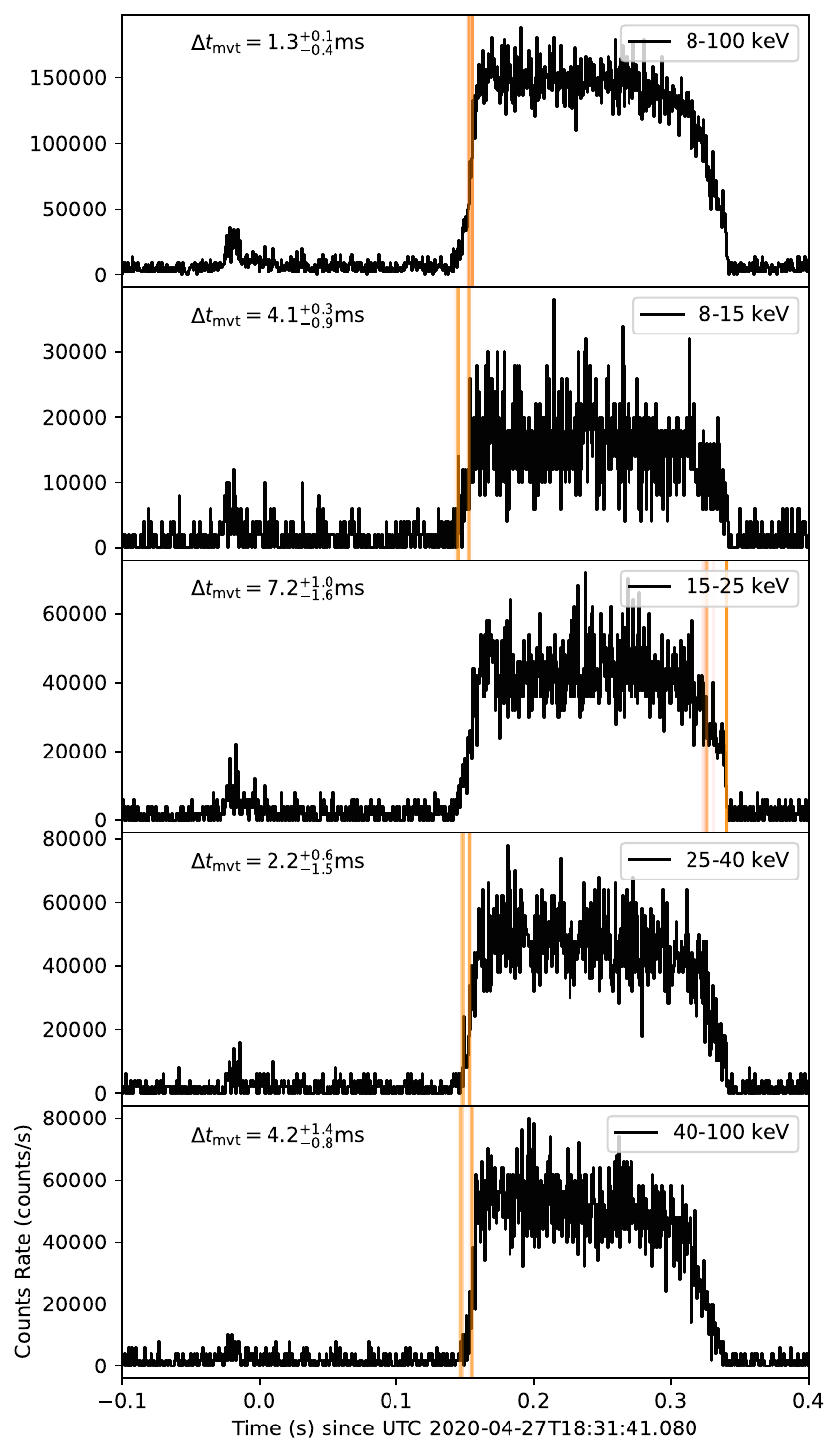}
\end{minipage}
\caption{The 0.5 ms binned light curves and MVTs of an burst ($T_0$=UTC 2020-04-27T18:31:41.080) from SGR J1935+2154 observed by the GBM NaI \# 0, 1, 3, 6, 7 and 9 in four energy bands. The MVT for different energy ranges is consistent within the error, except that 8-100 keV has a higher signal-to-noise ratio and therefore a smaller MVT (see Fig.~\ref{200427_fit}). The orange vertical lines are the locations of the two change points where the MVT was obtained.}\label{200427}
\end{figure}

\begin{figure}
\centering
\includegraphics[width=\columnwidth]{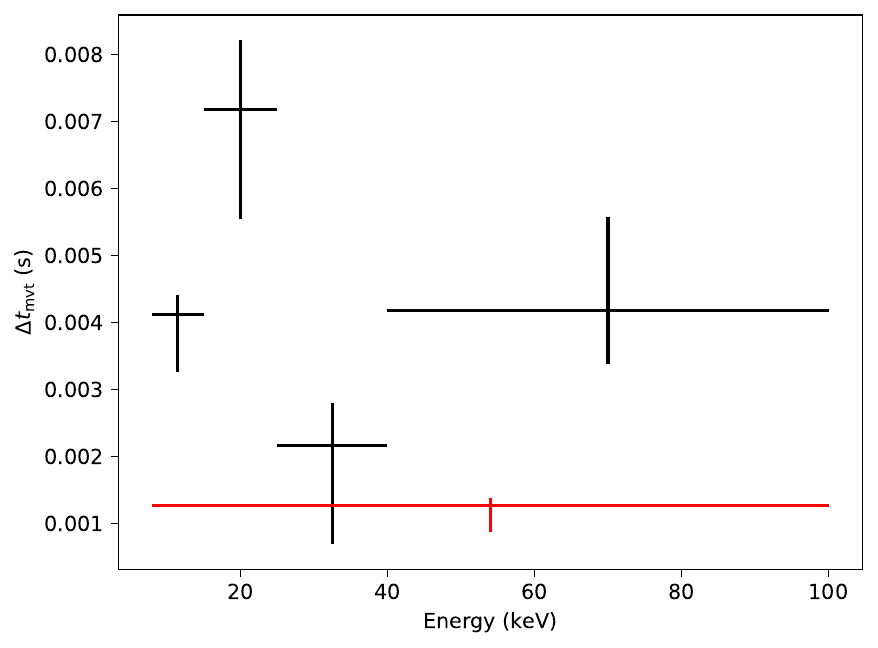}
\caption{The MVT vs. energy band for the burst UTC 2020-04-27T18:31:41.080. The light curve in 8-100 keV has a higher signal-to-noise ratio and therefore a smaller MVT (red error bar).}\label{200427_fit}
\end{figure}

\begin{figure}
\centering
\includegraphics[width=\columnwidth]{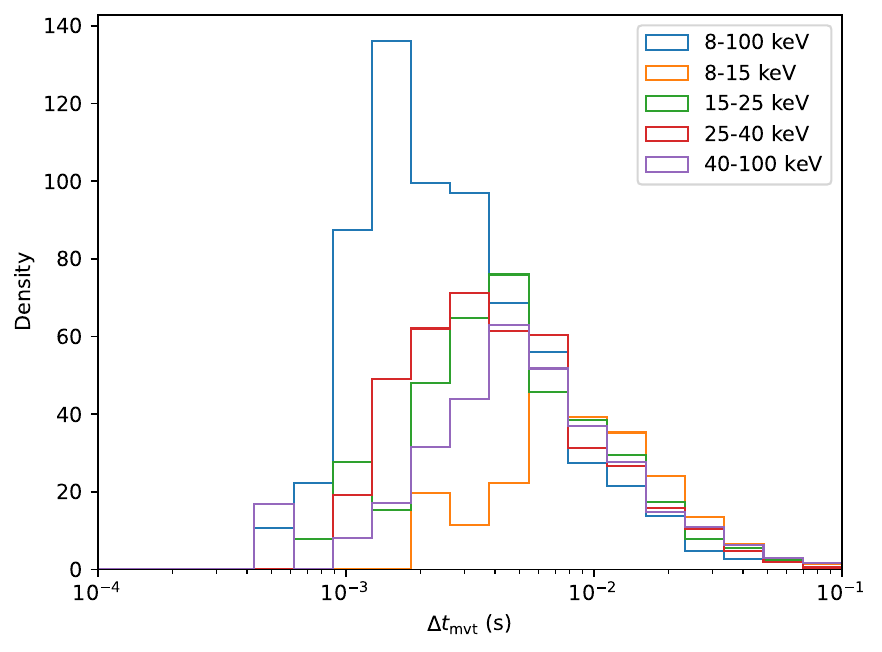}
\caption{Distribution of the MVTs ($\Delta t_{\rm mvt}$) for bursts from SGR J1935+2154 in 8-100 keV, 8-15 keV, 15-25 keV, 25-40 keV and 40-100 keV observed by GBM, respectively.}\label{fenbu_ee_gbm}
\end{figure}

\begin{figure*}
\centering
\begin{minipage}[t]{0.46\textwidth}
\centering
\includegraphics[width=\columnwidth]{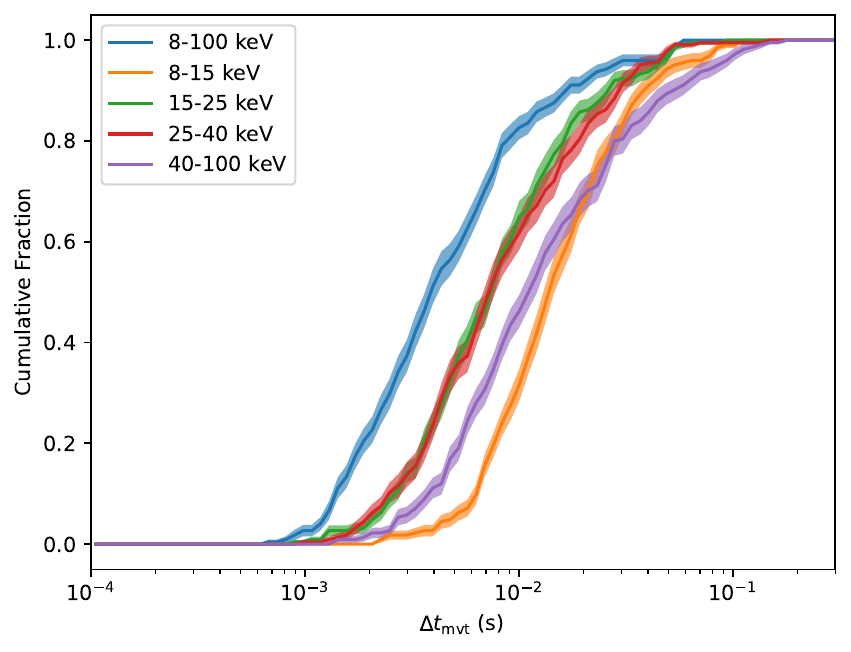}
\end{minipage}
\begin{minipage}[t]{0.46\textwidth}
\centering
\includegraphics[width=\columnwidth]{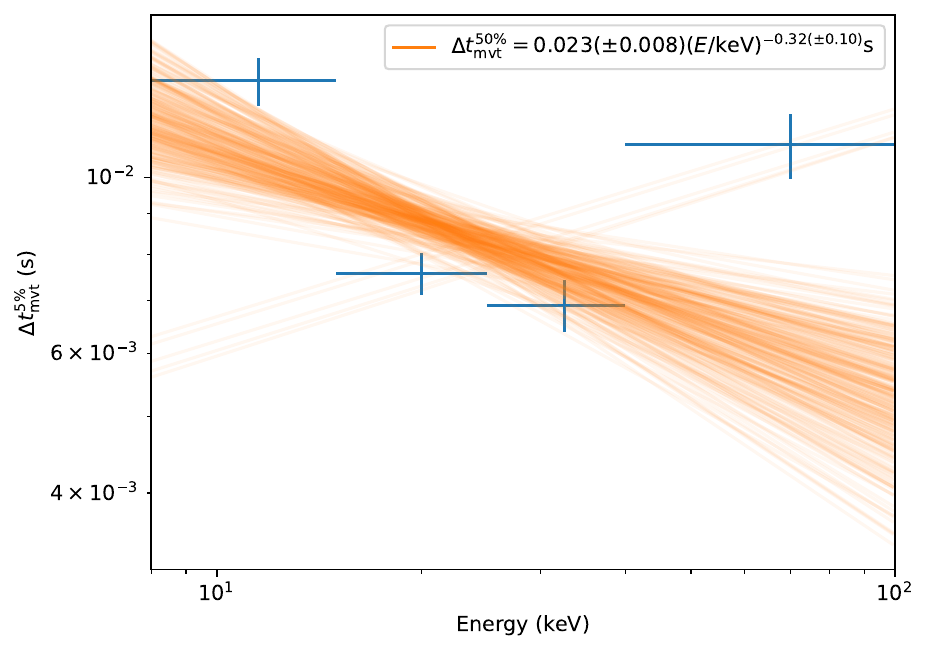}
\end{minipage}
\caption{Left panel: the cumulative curves of MVTs for different energy bands and 1$\sigma$ ranges. Right panel: the 50\% percentile of MVTs vs. energy band; the fitted yellow line is obtained by MCMC.}\label{tmins_e}
\end{figure*}

\begin{figure}
\centering
\includegraphics[width=\columnwidth]{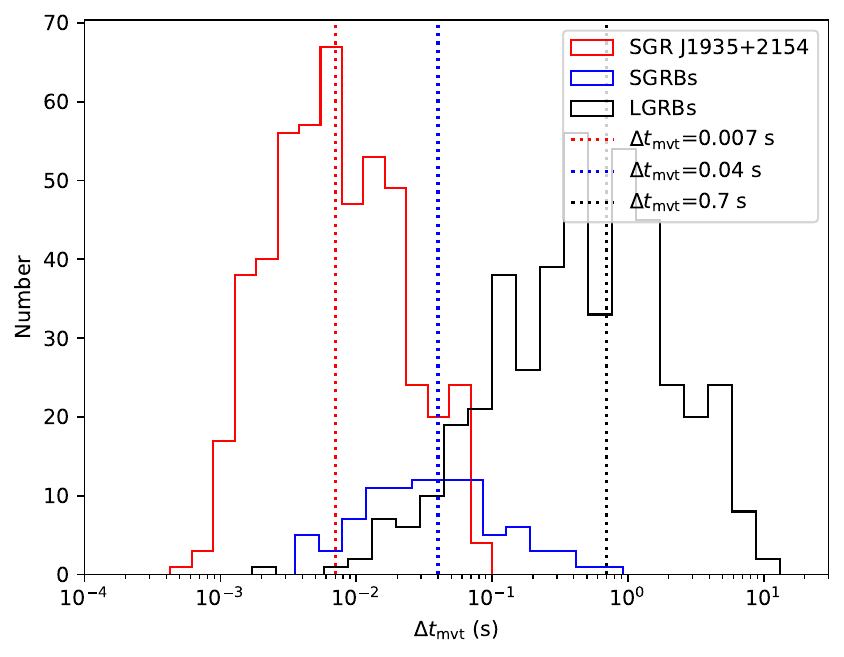}
\caption{Distribution of the MVTs ($\Delta t_{\rm mvt}$) for bursts from SGR J1935+2154, short and long GRBs. The results of GRBs are from \cite{golkhou2015energy}.}\label{3sources}
\end{figure}

\begin{figure}
\centering
\includegraphics[width=\columnwidth]{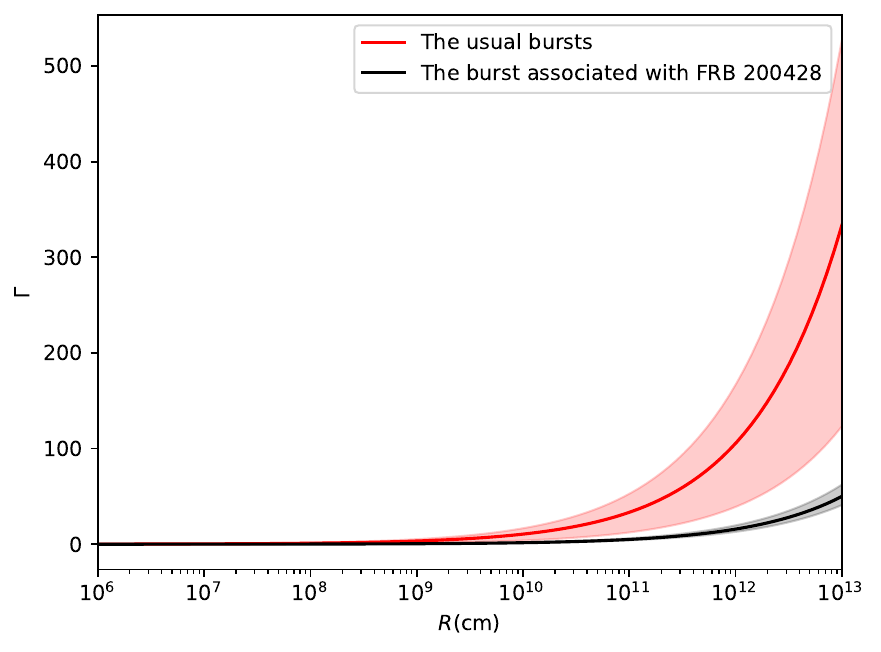}
\caption{The Lorentz factor of jet and the shock radius are constrained by MVT for the usual bursts and the burst associated with FRB 200428 based on the GRB-like model, respectively. The shaded area is the 1 $\sigma$ confidence interval. }\label{R_GAMMA}
\end{figure}

\section{Result}

\subsection{Distribution of MVTs}
Fig.~\ref{tmins} shows that the distribution of the MVTs of the bursts from SGR J1935+2154 peaks at $\sim$ 1.5 ms. The bursts observed by the three different satellites are not the same, but the distributions observed by the three satellites are consistent, with the MVT means of 23.4 ms, 17.0 ms and 14.9 ms observed by HXMT, GECAM and GBM, respectively. The standard deviations of MVTs observed by HXMT, GECAM and GBM are 33.1 ms, 21.3 ms and 22.3 ms, respectively. The medians of MVTs observed by HXMT, GECAM and GBM are 10.1 ms, 8.9 ms and 7.2 ms, respectively. Besides, a further K-S test comparing the populations observed by GBM and GECAM, GBM and HXMT gives the p-values of 0.33 and 0.02, respectively, supporting the null hypothesis that the two populations are drawn from the same underlying distribution.

\subsection{Consistency check of measurement results from different satellites}
We investigate whether the MVTs obtained by GBM and GECAM are consistent for the same burst. As shown in the left panel of Fig.~\ref{gbm_gecam}, their results are consistent within the error range, but more GECAM MVTs are larger than GBM MVTs. Although GBM and GECAM have similar energy responses \citep{2021ApJ...920...43X}, and the same deposition energy range is chosen to mitigate the effects of the differences between the two instruments, the measured MVT is very dependent on the strength of the signal \citep{golkhou2014uncovering}. In particular, these four bursts (UTCs 2021-09-12T16:52:07.950, 2021-09-18T22:58:52.150, 2022-01-05T07:06:40.800 and 2022-01-12T01:03:46.900) have a significantly lower signal-to-noise ratio in GECAM than in GBM, resulting in shorter pulses that are not observed by the former. The right panel of Fig.~\ref{gbm_gecam} shows that the error in MVT has a positive correlation with MVT.

Only one burst is jointly observed by GBM and HXMT, which is Trigger UTC 2020-05-16T18:12:52.120. The MVTs are $18.6_{-0.4}^{+1.5}$ ms and $2.1_{-0.2}^{+0.4}$ ms observed by HXMT and GBM, respectively. Because of the higher sensitivity of HXMT, resulting in a significantly higher signal-to-noise ratio for this burst on the HXMT than the GBM and the MVT observed by former is significantly smaller.

\subsection{The duration dependence of MVT}
The red, yellow and purple dots in the left panel of Fig.~\ref{3sources_mvt_t} show the relationship between MVT in 8-100 keV and duration (i.e. $T_{\rm bb}$) for the bursts from SGR J1935+2154. The Pearson correlation coefficients between MVT and duration observed by HXMT, GECAM and GBM are respectively 0.31, -0.18 and -0.04, with the Pearson p-values of 0.05, 0.10 and 0.34. The marginal evidence indicates that the MVT and duration are not correlated, similar to GRBs \citep{golkhou2014uncovering,maclachlan2012minimum}.

Because the burst ($T_0$=UTC 2020-04-28T14:34:24.150) associated with FRB 200428 saturated HXMT/HE \citep{li2021hxmt}, thus we calculate the MVT for 15-25 keV observed by HXMT/ME (see Fig.~\ref{FRB_LC}), which is $66.8^{+33.9}_{-23.4}$ ms. The right panel of Fig.~\ref{3sources_mvt_t} shows its comparison with MVTs of other bursts in 15-25 keV and the value is significantly larger than the MVTs of most bursts.

\subsection{The energy dependence of MVT}
In a previous study for GRBs, both their duration and pulse width are found to be energy dependent (e.g. \citealp{fenimore1995gamma}; \citealp{norris1996attributes}; \citealp{virgili2012spectral}; \citealp{bromberg2013short}; \citealp{kocevski2003connection}). For example, \cite{golkhou2015energy} reported two power-law relationships $\Delta t_{\rm min}^{50\%}=0.20(E/89\ {\rm keV})^{-0.53\pm 0.06}\ {\rm s}$ and $\Delta t_{\rm min}^{10\%}=0.01(E/48\ {\rm keV})^{-0.97\pm 0.20}\ {\rm s}$ between Kaplan–Meier median and 10\% values of MVTs and energy in GRBs, respectively, which is consistent with the relationship between the average width of pulses and the energy (\citealp{fenimore1995gamma}; \citealp{norris1996attributes}). To investigate whether the bursts from SGR J1935+2154 also follow a similar relationship, we therefore measure MVTs in four different energy bands (8-15, 15-25, 25-40, 40-100 keV), respectively. The MVTs at different energies observed by GBM, GECAM and HXMT are listed in Tables \ref{GBM_MVT}, \ref{GECAM_MVT} and \ref{HXMT_MVT}, respectively. Some MVTs that cannot be constrained due to the low signal-to-noise ratio are denoted by "-". Note that GECAM is mostly unable to limit MVT in 8-15 keV due to instrumental effects resulting in few counts observed below 20 keV.

Fig.~\ref{200427} shows the light curves of a bright burst ($T_0$=UTC 2020-04-27T18:31:41.080) and the MVTs in different energy bands, the MVTs for different energy ranges are consistent within the error, except that the light curve in 8-100 keV has a higher signal-to-noise ratio and therefore a smaller MVT (see Fig.~\ref{200427_fit}). Interestingly, the duration of the latter large pulse (from $\sim$ T0+0.14 s to T0+0.34 s) is also not obviously changed in the four energy ranges, which is different from GRBs where the timescales of the pulse width at energy bands usually are significantly different (\citealp{fenimore1995gamma}; \citealp{norris1996attributes}). 

To statistically investigate the relationship between MVT and energy, we show the distribution of the MVTs ($\Delta t_{\rm mvt}$) for bursts from SGR J1935+2154 in 8-100 keV, 8-15 keV, 15-25 keV, 25-40 keV and 40-100 keV observed by GBM, respectively. As shown in Fig.~\ref{fenbu_ee_gbm}, since the signal-to-noise ratio is obviously higher for 8-100 keV, it usually has a smaller MVT, however the peaks of the MVT distribution for the other four different energy ranges are generally consistent.

On the other hand, we also show the cumulative curves of different energy bands in Fig~\ref{tmins_e}; the uncertainties of the curves are obtained by Bootstrap method.
The 50\% percentile of MVTs versus energy band are also fitted by a power-law relationship $\Delta t_{\rm min}^{50\%}=0.023(\pm 0.008)(E/{\rm keV})^{-0.32(\pm 0.10)} {\rm s}$ (see Fig.\ref{200427_fit}), we note that the power-law index is consistent with the value of GRB ($\sim$ -0.53) within 2 $\sigma$ \citep{golkhou2015energy}. However, the rate of variation of MVT with energy is about an order of magnitude lower compared to GRBs and the index is consistent with 0 within the $\sim$ 3 $\sigma$ error range, that is, MVT may be independent of energy. It is worth noting here that since MVTs are very dependent on the signal-to-noise ratio, light curve in 8-100 keV energy band usually has the smallest MVT, while the MVT is larger due to the usually low signal-to-noise ratio of the burst from SGR J1935+2154 in 40-100 keV energy band, which may be the reason why the 50\% percentile of MVTs in 40-100 keV cannot fit it well. Therefore, we do not have sufficient evidence to claim that the MVT and energy of the bursts from SGR J1935+2154 are power-law dependent.

\subsection{Comparison between SGR J1935+2154 and GRBs}
Since GRB-like models are one class of the physical mechanisms responsible for FRBs from SGR J1935+2154 (e.g. \citealp{lyubarsky2014model,beloborodov2017flaring}; also see \citealp{2020Natur.587...45Z} for review), the similarities or differences between the SGR and GRBs may contain important information. We collect MVTs of SGRB and LGRB from \cite{golkhou2015energy} to compare with the values of the bursts from SGR J1935+2154. The MVT distributions in logarithmically uniform scale of SGR J1935+2154, SGRBs and LGRBs are shown in Fig.~\ref{3sources}, and their peak values are 7 ms, 40 ms and 700 ms, respectively. The medians of the MVTs are 7 ms, 40 ms and 480 ms, respectively. Therefore, the bursts from SGR J1935+2154 usually have smaller MVTs and more concentrated distribution than GRBs.

\section{Discussion and conclusion}
In this work, we perform a detailed MVT analysis for hundreds of bursts from SGR J1935+2154, including the MVT distributions observed by HXMT, GECAM and GBM, the relationship between MVT and energy as well as MVT and duration, which are compared with GRBs.

The MVTs of the X-ray bursts from SGR J1935+2154 peak at $\sim$ 2 ms, corresponding to a light travel time size 600 km, which supports that they originate from the magnetosphere in the pulsar-like models (see \citealp{2020Natur.587...45Z,2022arXiv221203972Z} for review). On the other hand, if the physical mechanisms of X-ray bursts from SGR J1935+2154 are the GRB-like models, we can constrain the relationship between the shock radius $R$ and the Lorentz factor $\Gamma$ of jet by $\Delta t_{\rm min}=R/(2\Gamma^2c)$. The Lorentz factor versus the shock radius is shown in Fig~\ref{R_GAMMA}. When the Lorentz factor is 100, the corresponding radius is $\sim 10^{12-13}$ cm, which is consistent with the GRB-like models (see \citealp{2020Natur.587...45Z,2022arXiv221203972Z} for review). It is worth noting here that the MVT of the burst associated with FRB 200428 is $66.8^{+33.9}_{-23.4}$ ms, which is significantly larger than most bursts and implies that its radiation mechanism is special.

Although the bursts observed by the three satellites are not the same, the MVT distributions are consistent. We check the consistency of the MVTs of the same bursts observed by the GBM and GECAM and find that although most are consistent within the error, the more GECAM MVTs are larger than GBM MVTs, most likely due to the instrumental effects of GECAM resulting in a weaker signal-to-noise ratio. Similarly, this effect also arises when studying the energy dependence of MVT; 
as shown in Fig.~\ref{tmins_e}, the MVTs of the most bursts at 40-100 keV are larger than those at 8-100 keV.

We do not find a statistically significant correlation between MVT and the duration of the bursts, which is similar to GRBs \citep{golkhou2014uncovering,maclachlan2012minimum}. In addition, we also investigate the possible energy dependence of MVT for the bursts, finding only a marginally significant (3 $\sigma$) power-law relationship $\Delta t_{\rm min}^{50\%}=0.023(\pm 0.008)(E/{\rm keV})^{-0.32(\pm 0.10)} {\rm s}$ between MVT and energy. In fact, according to this relationship, the MVTs at 10 keV and 60 keV are $\sim$ 11 ms and $\sim$ 6 ms, respectively. This is difficult to distinguish with the current signal-to-noise ratio, and a definitive answer may come from sufficient statistics observed by more advanced detectors (e.g. with larger effective area and better temporal resolution) and higher detection signal-to-noise ratios, or a joined analysis using the light curves obtained by multiple detectors.

We compare the MVTs of the burst from SGR J1935+2154 and GRBs. The median of MVTs for SGR J1935+2154 is 7 ms, which is significantly shorter than the 40 ms and 480 ms for the SGRBs or LGRBs, respectively. Besides, although there may be an energy dependence of MVT ($\sim$ 3$\sigma$ significance) with a power law relationship like GRBs, but the rate of variation at least is about an order of magnitude smaller.
The MVT distributions of the magnetar bursts and GRBs are also different, which can be used to identify the origin of the bursts. For example, GRB 200415A with magnetar origin has a MVT of $\sim$ 2 ms \citep{yang2020grb}, which is perfect consistent with that of SGR J1935+2154 rather than most GRBs. Therefore, if the MVT of a burst is only a few milliseconds, we should be pay attention to its possible magnetar origin.

\section*{Acknowledgments}
Thanks to the reviewer for his important suggestions on the MVT calculation method. This work is supported by the National Key R\&D Program of China (2022YFF0711404, 2021YFA0718500), the Scientific Research Project of the Guizhou Provincial Education (Nos. KY[2022]123, KY[2022]132, KY[2022]137), and partially supported by International Partnership Program of Chinese Academy of Sciences (Grant No. 113111KYSB20190020). %% HXMT research
The authors also thank supports from 
the Strategic Priority Research Program on Space Science, the Chinese Academy of Sciences (Grant No.
XDA15010100, %% Peng wenxi
XDA15360100, XDA15360102, %% GECAM satellite and payload
XDA15360300, %% GECAM science appliacation
XDA15052700), %% Xiong shaolin, GECAM data analysis
 the National Natural Science Foundation of China (Projects: 12061131007 %% ?
 and Grant No. 12173038), the Foundation of Education Bureau of Guizhou Province, China (Grant No. KY (2020) 003), Guizhou Provincial Science and Technology Foundation (Nos. ZK[2022]304, ZK[2022]322), the Major Science and Technology Program of Xinjiang Uygur Autonomous Region (No.2022A03013-4), the Joint Research Fund in Astronomy (Grant Nos. U1931101) under
cooperative agreement between the National Natural Science Foundation of China (NSFC) and Chinese Academy of Sciences (CAS). %%Li Xinqiao, GECAM space environment 
 S. Xiao is grateful to W. Xiao, G. Q. Wang, J. H. Li and Q. Q. Xiao for their useful comments. 

\newpage

\bibliography{main}

\begin{thebibliography}{}
\expandafter\ifx\csname natexlab\endcsname\relax\def\natexlab#1{#1}\fi
\providecommand{\url}[1]{\href{#1}{#1}}
\providecommand{\dodoi}[1]{doi:~\href{http://doi.org/#1}{\nolinkurl{#1}}}
\providecommand{\doeprint}[1]{\href{http://ascl.net/#1}{\nolinkurl{http://ascl.net/#1}}}
\providecommand{\doarXiv}[1]{\href{https://arxiv.org/abs/#1}{\nolinkurl{https://arxiv.org/abs/#1}}}

\bibitem[{Abbott {et~al.}(2017)Abbott, Abbott, Abbott, Acernese, Ackley, Adams,
  Adams, Addesso, Adhikari, Adya, {et~al.}}]{abbott2017gravitational}
Abbott, B.~P., Abbott, R., Abbott, T., {et~al.} 2017, The Astrophysical Journal
  Letters, 848, L13

\bibitem[{Ackermann {et~al.}(2014)Ackermann, Ajello, Asano, Atwood, Axelsson,
  Baldini, Ballet, Barbiellini, Baring, Bastieri,
  {et~al.}}]{ackermann2014fermi}
Ackermann, M., Ajello, M., Asano, K., {et~al.} 2014, Science, 343, 42

\bibitem[{Beloborodov(2017)}]{beloborodov2017flaring}
Beloborodov, A.~M. 2017, The Astrophysical Journal Letters, 843, L26

\bibitem[{Bochenek {et~al.}(2020)Bochenek, Ravi, Belov, Hallinan, Kocz,
  Kulkarni, \& McKenna}]{bochenek2020fast}
Bochenek, C.~D., Ravi, V., Belov, K.~V., {et~al.} 2020, Nature, 587, 59

\bibitem[{Briggs {et~al.}(2013)Briggs, Xiong, Connaughton, Tierney,
  Fitzpatrick, Foley, Grove, Chekhtman, Gibby, Fishman,
  {et~al.}}]{briggs2013terrestrial}
Briggs, M.~S., Xiong, S., Connaughton, V., {et~al.} 2013, Journal of
  Geophysical Research: Space Physics, 118, 3805

\bibitem[{Bromberg {et~al.}(2013)Bromberg, Nakar, Piran,
  {et~al.}}]{bromberg2013short}
Bromberg, O., Nakar, E., Piran, T., {et~al.} 2013, The Astrophysical Journal,
  764, 179

\bibitem[{Cai {et~al.}(2022)Cai, Xue, Li, Xiong, Zhang, Lin, Li, Ge, Zhao,
  Song, {et~al.}}]{cai2022insight}
Cai, C., Xue, W.-C., Li, C.-K., {et~al.} 2022, The Astrophysical Journal
  Supplement Series, 260, 24

\bibitem[{{Camisasca} {et~al.}(2023){Camisasca}, {Guidorzi}, {Amati},
  {Frontera}, {Song}, {Xiao}, {Xiong}, {Zhang}, {Margutti}, {Kobayashi},
  {Mundell}, {Ge}, {Gomboc}, {Jia}, {Jordana-Mitjans}, {Li}, {Li}, {Maccary},
  {Shrestha}, {Xue}, \& {Zhang}}]{2023A&A...671A.112C}
{Camisasca}, A.~E., {Guidorzi}, C., {Amati}, L., {et~al.} 2023, \aap, 671,
  A112, \dodoi{10.1051/0004-6361/202245657}

\bibitem[{Cao {et~al.}(2020)Cao, Jiang, Meng, Zhang, Luo, Yang, Zhang, Gu, Sun,
  Liu, {et~al.}}]{cao2020medium}
Cao, X., Jiang, W., Meng, B., {et~al.} 2020, SCIENCE CHINA Physics, Mechanics
  \& Astronomy, 63, 1

\bibitem[{Chen {et~al.}(2020)Chen, Cui, Li, Wang, Xu, Lu, Wang, Chen, Han, Hu,
  {et~al.}}]{chen2020low}
Chen, Y., Cui, W., Li, W., {et~al.} 2020, SCIENCE CHINA Physics, Mechanics \&
  Astronomy, 63, 1

\bibitem[{Fenimore {et~al.}(1993)Fenimore, Epstein, \& Ho}]{fenimore1993escape}
Fenimore, E., Epstein, R., \& Ho, C. 1993, Astronomy and Astrophysics
  Supplement Series, 97, 59

\bibitem[{Fenimore {et~al.}(1995)Fenimore, Norris, Bonnell, Nemiroff,
  {et~al.}}]{fenimore1995gamma}
Fenimore, E., Norris, J., Bonnell, J., Nemiroff, R., {et~al.} 1995, The
  Astrophysical Journal, 448, L101

\bibitem[{Golkhou \& Butler(2014)}]{golkhou2014uncovering}
Golkhou, V.~Z., \& Butler, N.~R. 2014, The Astrophysical Journal, 787, 90

\bibitem[{Golkhou {et~al.}(2015)Golkhou, Butler, \&
  Littlejohns}]{golkhou2015energy}
Golkhou, V.~Z., Butler, N.~R., \& Littlejohns, O.~M. 2015, The Astrophysical
  Journal, 811, 93

\bibitem[{Guidorzi {et~al.}(2016)Guidorzi, Dichiara, \&
  Amati}]{guidorzi2016individual}
Guidorzi, C., Dichiara, S., \& Amati, L. 2016, Astronomy \& Astrophysics, 589,
  A98

\bibitem[{Hurley {et~al.}(1999)Hurley, Cline, Mazets, Barthelmy, Butterworth,
  Marshall, Palmer, Aptekar, Golenetskii, Il'Inskii,
  {et~al.}}]{hurley1999giant}
Hurley, K., Cline, T., Mazets, E., {et~al.} 1999, Nature, 397, 41

\bibitem[{Israel {et~al.}(2016)Israel, Esposito, Rea, Coti~Zelati, Tiengo,
  Campana, Mereghetti, Rodr{\'\i}guez~Castillo, G{\"o}tz, Burgay,
  {et~al.}}]{israel2016discovery}
Israel, G.~L., Esposito, P., Rea, N., {et~al.} 2016, Monthly Notices of the
  Royal Astronomical Society, 457, 3448

\bibitem[{Kocevski \& Liang(2003)}]{kocevski2003connection}
Kocevski, D., \& Liang, E. 2003, The Astrophysical Journal, 594, 385

\bibitem[{Li {et~al.}(2021)Li, Lin, Xiong, Ge, Li, Li, Lu, Zhang, Tuo, Nang,
  {et~al.}}]{li2021hxmt}
Li, C., Lin, L., Xiong, S., {et~al.} 2021, Nature Astronomy, 5, 378

\bibitem[{Lin {et~al.}(2020)Lin, G{\"o}{\u{g}}{\"u}{\c{s}}, Roberts, Baring,
  Kouveliotou, Kaneko, Van~der Horst, \& Younes}]{lin2020fermi}
Lin, L., G{\"o}{\u{g}}{\"u}{\c{s}}, E., Roberts, O.~J., {et~al.} 2020, The
  Astrophysical Journal Letters, 902, L43

\bibitem[{Liu {et~al.}(2020)Liu, Zhang, Li, Lu, Chang, Li, Zhang, Jin, Yu,
  Zhang, {et~al.}}]{liu2020high}
Liu, C., Zhang, Y., Li, X., {et~al.} 2020, SCIENCE CHINA Physics, Mechanics \&
  Astronomy, 63, 1

\bibitem[{Liu {et~al.}(2021)Liu, Gong, Li, Wen, An, Cai, Chang, Chen, Chen, Du,
  {et~al.}}]{liu2021sipm}
Liu, Y., Gong, K., Li, X., {et~al.} 2021, arXiv preprint arXiv:2112.04786

\bibitem[{Lyubarsky(2014)}]{lyubarsky2014model}
Lyubarsky, Y. 2014, Monthly Notices of the Royal Astronomical Society: Letters,
  442, L9

\bibitem[{MacLachlan {et~al.}(2012)MacLachlan, Shenoy, Sonbas, Dhuga,
  Eskandarian, Maximon, \& Parke}]{maclachlan2012minimum}
MacLachlan, G., Shenoy, A., Sonbas, E., {et~al.} 2012, Monthly Notices of the
  Royal Astronomical Society: Letters, 425, L32

\bibitem[{{MacLachlan} {et~al.}(2013){MacLachlan}, {Shenoy}, {Sonbas}, {Dhuga},
  {Cobb}, {Ukwatta}, {Morris}, {Eskandarian}, {Maximon}, \&
  {Parke}}]{2013MNRAS.432..857M}
{MacLachlan}, G.~A., {Shenoy}, A., {Sonbas}, E., {et~al.} 2013, \mnras, 432,
  857, \dodoi{10.1093/mnras/stt241}

\bibitem[{Meegan {et~al.}(2009)Meegan, Lichti, Bhat, Bissaldi, Briggs,
  Connaughton, Diehl, Fishman, Greiner, Hoover, {et~al.}}]{meegan2009fermi}
Meegan, C., Lichti, G., Bhat, P., {et~al.} 2009, The Astrophysical Journal,
  702, 791

\bibitem[{Mereghetti {et~al.}(2020)Mereghetti, Savchenko, Ferrigno, G{\"o}tz,
  Rigoselli, Tiengo, Bazzano, Bozzo, Coleiro, Courvoisier,
  {et~al.}}]{mereghetti2020integral}
Mereghetti, S., Savchenko, V., Ferrigno, C., {et~al.} 2020, The Astrophysical
  Journal Letters, 898, L29

\bibitem[{Norris {et~al.}(1996)Norris, Nemiroff, Bonnell, Scargle, Kouveliotou,
  Paciesas, Meegan, \& Fishman}]{norris1996attributes}
Norris, J., Nemiroff, R., Bonnell, J., {et~al.} 1996, The Astrophysical
  Journal, 459, 393

\bibitem[{Ridnaia {et~al.}(2021)Ridnaia, Svinkin, Frederiks, Bykov, Popov,
  Aptekar, Golenetskii, Lysenko, Tsvetkova, Ulanov,
  {et~al.}}]{ridnaia2021peculiar}
Ridnaia, A., Svinkin, D., Frederiks, D., {et~al.} 2021, Nature Astronomy, 5,
  372

\bibitem[{Scargle {et~al.}(2013)Scargle, Norris, Jackson, \&
  Chiang}]{scargle2013studies}
Scargle, J.~D., Norris, J.~P., Jackson, B., \& Chiang, J. 2013, The
  Astrophysical Journal, 764, 167

\bibitem[{{Schmidt}(1978)}]{1978Natur.271..525S}
{Schmidt}, W.~K.~H. 1978, \nat, 271, 525, \dodoi{10.1038/271525a0}

\bibitem[{Svinkin {et~al.}(2021)Svinkin, Frederiks, Hurley, Aptekar,
  Golenetskii, Lysenko, Ridnaia, Tsvetkova, Ulanov, Cline,
  {et~al.}}]{svinkin2021bright}
Svinkin, D., Frederiks, D., Hurley, K., {et~al.} 2021, Nature, 589, 211

\bibitem[{Tavani {et~al.}(2021)Tavani, Casentini, Ursi, Verrecchia, Addis,
  Antonelli, Argan, Barbiellini, Baroncelli, Bernardi, {et~al.}}]{tavani2021x}
Tavani, M., Casentini, C., Ursi, A., {et~al.} 2021, Nature Astronomy, 5, 401

\bibitem[{{Titarchuk} {et~al.}(2007){Titarchuk}, {Shaposhnikov}, \&
  {Arefiev}}]{2007ApJ...660..556T}
{Titarchuk}, L., {Shaposhnikov}, N., \& {Arefiev}, V. 2007, \apj, 660, 556,
  \dodoi{10.1086/512027}

\bibitem[{Vianello {et~al.}(2018)Vianello, Gill, Granot, Omodei, Cohen-Tanugi,
  \& Longo}]{vianello2018bright}
Vianello, G., Gill, R., Granot, J., {et~al.} 2018, The Astrophysical Journal,
  864, 163

\bibitem[{Virgili {et~al.}(2012)Virgili, Qin, Zhang, \&
  Liang}]{virgili2012spectral}
Virgili, F.~J., Qin, Y., Zhang, B., \& Liang, E. 2012, Monthly Notices of the
  Royal Astronomical Society, 424, 2821

\bibitem[{{Woods} \& {Thompson}(2006)}]{2006csxs.book..547W}
{Woods}, P.~M., \& {Thompson}, C. 2006, in Compact stellar X-ray sources,
  Vol.~39, 547--586, \dodoi{10.48550/arXiv.astro-ph/0406133}

\bibitem[{Woosley \& Bloom(2006)}]{woosley2006supernova}
Woosley, S., \& Bloom, J. 2006, Annu. Rev. Astron. Astrophys., 44, 507

\bibitem[{Xiao {et~al.}(2020)Xiao, Xiong, Liu, Li, Zhang, Ge, Cai, Yi, Zhu,
  Chen, {et~al.}}]{xiao2020deadtime}
Xiao, S., Xiong, S., Liu, C., {et~al.} 2020, Journal of High Energy
  Astrophysics, 26, 58

\bibitem[{{Xiao} {et~al.}(2021){Xiao}, {Xiong}, {Zhang}, {Song}, {Lu}, {Huang},
  {Cai}, {Yi}, {Song}, {Chen}, {Ge}, {Liu}, {Li}, {Li}, \&
  {Zhao}}]{2021ApJ...920...43X}
{Xiao}, S., {Xiong}, S.~L., {Zhang}, S.~N., {et~al.} 2021, \apj, 920, 43,
  \dodoi{10.3847/1538-4357/ac1420}

\bibitem[{Xiao {et~al.}(2022{\natexlab{a}})Xiao, Zhang, Zhu, Xiong, Gao, Xu,
  Zhang, Peng, Li, Zhang, {et~al.}}]{xiao2022quasi}
Xiao, S., Zhang, Y.-Q., Zhu, Z.-P., {et~al.} 2022{\natexlab{a}}, arXiv preprint
  arXiv:2205.02186

\bibitem[{Xiao {et~al.}(2022{\natexlab{b}})Xiao, Xiong, Cai, Song, Zheng, Peng,
  Wang, Qiao, Guo, Wang, {et~al.}}]{xiao2022energetic}
Xiao, S., Xiong, S.-L., Cai, C., {et~al.} 2022{\natexlab{b}}, Monthly Notices
  of the Royal Astronomical Society, 514, 2397

\bibitem[{Xiao {et~al.}(2022{\natexlab{c}})Xiao, Liu, Peng, An, Xiong, Tuo,
  Gong, Zhang, Zhang, Zheng, {et~al.}}]{xiao2022ground}
Xiao, S., Liu, Y., Peng, W., {et~al.} 2022{\natexlab{c}}, Monthly Notices of
  the Royal Astronomical Society, 511, 964

\bibitem[{{Xiao} {et~al.}(2023){Xiao}, {Tuo}, {Zhang}, {Xiong}, {Lin}, {Zhang},
  {Wang}, {Xue}, {Cai}, {Gao}, {Li}, {Li}, {Zheng}, {Liu}, {Wang}, {Wang},
  {Peng}, {Liu}, {Li}, {Wen}, {An}, {Song}, {Zheng}, {Zhang}, {Dong}, {Xie},
  {Feng}, {Ma}, {Wang}, {Luo}, {Dang}, {Shang}, {Zhi}, \&
  {Li}}]{xiao2023discovery}
{Xiao}, S., {Tuo}, Y.-L., {Zhang}, S.-N., {et~al.} 2023, \mnras, 521, 5308,
  \dodoi{10.1093/mnras/stad885}

\bibitem[{Xie {et~al.}(2022)Xie, Cai, Xiong, Yu, Zhang, Lin, Zhang, Xue, Liu,
  Zhao, {et~al.}}]{xie2022revisit}
Xie, S.-L., Cai, C., Xiong, S.-L., {et~al.} 2022, Monthly Notices of the Royal
  Astronomical Society, 517, 3854

\bibitem[{Yang {et~al.}(2020)Yang, Chand, Zhang, Yang, Zou, Yang, Zhao, Shao,
  Xiong, Luo, {et~al.}}]{yang2020grb}
Yang, J., Chand, V., Zhang, B.-B., {et~al.} 2020, The Astrophysical Journal,
  899, 106

\bibitem[{Younes {et~al.}(2021)Younes, Baring, Kouveliotou, Arzoumanian, Enoto,
  Doty, Gendreau, G{\"o}{\u{g}}{\"u}{\c{s}}, Guillot, G{\"u}ver,
  {et~al.}}]{younes2021broadband}
Younes, G., Baring, M., Kouveliotou, C., {et~al.} 2021, Nature Astronomy, 5,
  408

\bibitem[{{Zhang}(2020)}]{2020Natur.587...45Z}
{Zhang}, B. 2020, \nat, 587, 45, \dodoi{10.1038/s41586-020-2828-1}

\bibitem[{Zhang(2022)}]{2022arXiv221203972Z}
Zhang, B. 2022, arXiv preprint arXiv:2212.03972

\bibitem[{Zhang {et~al.}(2020)Zhang, Li, Lu, Song, Xu, Liu, Chen, Cao, Bu,
  Chang, {et~al.}}]{zhang2020overview}
Zhang, S.-N., Li, T., Lu, F., {et~al.} 2020, SCIENCE CHINA Physics, Mechanics
  \& Astronomy, 63, 1

\bibitem[{Zou {et~al.}(2021)Zou, Zhang, Zhang, Yang, Shao, \&
  Wang}]{zou2021periodicity}
Zou, J.-H., Zhang, B.-B., Zhang, G.-Q., {et~al.} 2021, The Astrophysical
  Journal Letters, 923, L30

\end{thebibliography}

% \onecolumn
\newpage
\setlength{\tabcolsep}{1.2mm}
\begin{longtable*}{cccccccc}
\endfirsthead
\hline
\endhead
\hline
\endfoot
\hline
\endlastfoot\\
\caption{MVTs in different energy ranges for X-ray bursts from SGR J1935+2154 observed by Fermi/GBM.}\label{GBM_MVT}\\
\hline
Trigger time (UTC)& $T_{bb}$ & $\Delta t_{\rm mvt}$ (8-100 keV)& $\Delta t_{\rm mvt}$ (8-15 keV)& $\Delta t_{\rm mvt}$ (15-25 keV)&$\Delta t_{\rm mvt}$ (25-40 keV) & $\Delta t_{\rm mvt}$ (40-100 keV)&\\

\hline
2014-07-05T09:32:48.640 & 0.088 & $0.0163_{-0.0059}^{+0.0019}$ & -                            & $0.0471_{-0.0100}^{+0.0016}$ & $0.0517_{-0.0032}^{+0.0124}$ & $0.0438_{-0.0053}^{+0.0038}$ \\
2014-07-05T09:37:34.480 & 0.020 & $0.0042_{-0.0006}^{+0.0012}$ & $0.0102_{-0.0015}^{+0.0056}$ & $0.0040_{-0.0003}^{+0.0004}$ & $0.0097_{-0.0003}^{+0.0015}$ & $0.0089_{-0.0018}^{+0.0007}$ \\
2015-02-22T13:33:00.000 & 0.091 & $0.0048_{-0.0048}^{+0.0034}$ & -                            & $0.0459_{-0.0890}^{+0.0052}$ & -                            & $0.0407_{-0.0214}^{+0.0158}$ \\
2015-02-22T14:33:20.240 & 0.252 & $0.0114_{-0.0009}^{+0.0008}$ & -                            & -                            & -                            & $0.1509_{-0.0133}^{+0.0055}$ \\
2015-02-22T17:57:05.960 & 0.065 & $0.0008_{-0.0004}^{+0.0008}$ & $0.0089_{-0.0019}^{+0.0002}$ & $0.0097_{-0.0025}^{+0.0008}$ & $0.0031_{-0.0007}^{+0.0051}$ & $0.0047_{-0.0012}^{+0.0033}$ \\
\hline
\end{longtable*}

\newpage
\begin{longtable*}{cccccccc}
\endfirsthead
\hline
\endhead
\hline
\endfoot
\hline
\endlastfoot\\
\caption{MVTs in different energy ranges for X-ray bursts from SGR J1935+2154 observed by GECAM-B.}\label{GECAM_MVT}\\
\hline
Trigger time (UTC)& $T_{bb}$ & $\Delta t_{\rm mvt}$ (8-100 keV)&$\Delta t_{\rm mvt}$ (8-15 keV) & $\Delta t_{\rm mvt}$ (15-25 keV)&$\Delta t_{\rm mvt}$ (25-40 keV) & $\Delta t_{\rm mvt}$ (40-100 keV)&\\

\hline
2021-01-27T06:50:20.750 & 0.324 & $0.0007_{-0.0002}^{+0.0002}$ & $0.0369_{-0.0009}^{+0.0031}$ & $0.0043_{-0.0018}^{+0.0005}$ & $0.0045_{-0.0003}^{+0.0003}$ & $0.0078_{-0.0009}^{+0.0003}$ \\
2021-01-30T08:39:53.840 & 0.160 & $0.0802_{-0.0016}^{+0.0045}$ & $0.0803_{-0.0044}^{+0.0068}$ & $0.0827_{-0.0140}^{+0.0151}$ & $0.1001_{-0.0470}^{+0.0060}$ & $0.0830_{-0.0286}^{+0.0053}$ \\
2021-01-30T10:35:35.120 & 0.744 & $0.0089_{-0.0004}^{+0.0010}$ & $0.0469_{-0.0285}^{+0.0051}$ & $0.0095_{-0.0031}^{+0.0003}$ & -                            & -                            \\
2021-02-11T13:43:16.760 & 0.033 & $0.0165_{-0.0387}^{+0.0099}$ & -                            & -                            & -                            & -                            \\
2021-02-16T22:20:39.600 & 0.299 & $0.0013_{-0.0005}^{+0.0004}$ & $0.0114_{-0.0045}^{+0.0017}$ & $0.0049_{-0.0014}^{+0.0005}$ & $0.0157_{-0.0008}^{+0.0114}$ & $0.0040_{-0.0004}^{+0.0002}$ \\
\end{longtable*}

\newpage

\begin{longtable*}{cccccccc}
\endfirsthead
\hline
\endhead
\hline
\endfoot
\hline
\endlastfoot\\
\caption{MVTs in different energy ranges for X-ray bursts from SGR J1935+2154 observed by {\it Insight}-HXMT.}\label{HXMT_MVT}\\
\hline
Trigger time (UTC)& $T_{bb}$ & $\Delta t_{\rm mvt}$ (8-100 keV)& $\Delta t_{\rm mvt}$ (8-15 keV)& $\Delta t_{\rm mvt}$ (15-25 keV)&$\Delta t_{\rm mvt}$ (25-40 keV) & $\Delta t_{\rm mvt}$ (40-100 keV)&\\

\hline

2020-04-28T08:05:50.080 & 0.459 & $0.0062_{-0.0010}^{+0.0023}$ & $0.0512_{-0.0022}^{+0.0023}$ & -                            & $0.0213_{-0.0005}^{+0.0022}$ & -                            \\
2020-04-28T09:08:44.280 & 0.167 & $0.0083_{-0.0008}^{+0.0020}$ & $0.0071_{-0.5284}^{+0.0027}$ & $0.0111_{-0.0038}^{+0.0024}$ & $0.0074_{-0.0004}^{+0.0040}$ & $0.0066_{-0.0034}^{+0.0002}$ \\
2020-04-28T09:40:10.980 & 0.097 & $0.0483_{-0.0217}^{+0.0080}$ & -                            & -                            & -                            & -                            \\
2020-04-28T10:54:23.850 & 0.044 & $0.0085_{-0.0022}^{+0.0006}$ & $0.0198_{-0.0005}^{+0.0175}$ & -                            & -                            & -                            \\
2020-04-28T11:12:58.520 & 0.056 & $0.0060_{-0.0009}^{+0.0029}$ & -                            & -                            & $0.0207_{-0.0044}^{+0.0060}$ & -                            \\
\hline
\end{longtable*}

\end{document}